\documentclass[iop,revtex4]{emulateapj}

\begin{document}

\newcommand{\kms}{km~s$^{-1}$\,}
\newcommand{\msun}{$M_\odot$\,}

% for float placement:
\renewcommand{\topfraction}{1.0}
\renewcommand{\bottomfraction}{1.0}
\renewcommand{\textfraction}{0.0}

\title{Observations of red giants with suspected massive companions}
\shorttitle{Red giants with suspected  massive companions}

\author{Valeri V. Makarov}
\affil{ U.S. Naval Observatory, 3450 Massachusetts Ave., Washington, DC 20392-5420, USA}
\email{valeri.makarov@gmail.com}

\author{Andrei Tokovinin}
\affil{Cerro Tololo Inter-American Observatory, Casilla 603, La Serena, Chile}
\email{atokovinin@ctio.noao.edu}

\begin{abstract}
Motivated  by the  existence of  binary systems  where  a stellar-mass
black hole is bound to a normal star, we selected four red giants with
large radial velocity (RV) variation from the survey of Space Interferometry Mission (SIM) grid stars
and  monitored their  RVs  for  several months.   None  turned out  to
contain a massive companion above  2.5 solar masses. The red giant TYC
9299-1080-1  with  a  large  RV   and  a  large  proper  motion  is  a
single-lined spectroscopic binary with a  period of 81 days. It is an extreme
halo object moving at 350 km s$^{-1}$ almost directly toward the Galactic center.
HD 206092
is a double-lined binary with a  short period of 4.37 days. It belongs
to the rare class of active RS CVn-type binaries with evolved
primary  components, apparently  undergoing mass  transfer.  The X-ray
luminosity of  HD 206092 is about  twice as high as  the most luminous
coronal X-ray  emitters observed  by ROSAT, including  II Peg  and the
prototype  star  RS  CVn.   HD~318347  has  a  variable  double-peaked
emission-line spectrum  (not a giant), while HD~324668  has a constant
RV.  Despite the overall good quality of the SIM survey data confirmed
by a comparison with Gaia Data Release 2 (DR2) mean radial velocities, the few large RV  
variations are explained, mostly, by
erroneous data.   We discuss the significance of  the non-detection of
massive companions in the SIM grid sample and the associated work.
\end{abstract}

\keywords{
binaries --- spectroscopic}

%%----

%----------------------------------------------------------
\section{Introduction}
\label{sec:intro}

The  first detection of  gravitational wave  signal, GW150914,  by the
LIGO and Virgo  detectors incited a renewed interest  in the evolution
of  binary   systems  including  degenerate  stars   and  black  holes (BH)
\citep{Abbott2016}.  It showed that  stellar-mass black  holes (StMBH)
with  masses between  $2.5 M_\odot$  and  several tens  of solar  mass
should be present in binary  systems in significant numbers, unless we
were lucky  to observe an  exceedingly rare and fortuitous  event. The
detectable burst lasted for $\sim$0.2 s, but it probably took billions
of years for  the binary system to go through  the stages of evolution
leading  to this  event,  releasing approximately three  solar masses  of
energy  in  the form  of  gravitational radiation.  It  is  now up  to
observational astrophysics  to confirm this scenario  by finding StMBH
in Galactic binaries.

Black  holes  in  tight  binaries  with regular  star  companions  are
observable as  powerful and  variable X-ray sources.  There are  a few
hundred known  objects of  this type  in the Milky  Way, but  only two
dozen  have  been  dynamically  confirmed as  StMBH  \citep{cas}.  The
observable phenomena  are caused  not by the  StMBH itself but  by the
high-energy processes  in the  accreted material. The  orbital periods
range from less than  3 hours to a month, but most  are shorter than 1
day.  Most  of  these  tight  X-ray binaries  include  a  dwarf  donor
companion,  but a  few systems  with giant  companions are  also known
\citep{li}. The estimated masses of the BH companions are greater than
$\sim2.7$  $M_\odot$. Such  tight  systems  must be  the  result of  a
long-term   ($>10^9$ yr)   tidal  evolution   or   of  rare   dynamical
events.  However, it  is reasonable  to  expect that  the majority  of
Galactic  StMBH reside in  wider pairs  with normal  stars (separation
$<1$ au  for main sequence  companions and a  few astronomical units for  red giants),
where there  may be  no or very  little observable effects.  There are
three main possible ways to detect such binaries.

\begin{itemize}
\item
Precision  astrometry can  reveal  the reflex  orbital  motion of  the
stellar companion  around the barycenter  of the system. Based  on the
theoretical   expectations   of   the   rate  of   failed   supernovae
\citep[e.g.][]{Woosley1986},   \citet{gou}
predicted that $\sim$30 binaries containing StMBH remnants are present
in  the Hipparcos  catalog,  but none  has  been found,  owing to  the
limited  sensitivity.  Some candidate  binaries have  been reprocessed
\citep{Goldin2006,Goldin2007},  but  the   large  uncertainty  of  the
orbital parameters precluded definite detection of StMBHs. The renewed
interest in this  direction is focused now on  the {\it Gaia} mission, which
is expected to discover StMBH in scores \citep{Mashian2017,Kinugawa2018}. 

\item
Radial  velocity (RV)  variations can  reveal  large-amplitude orbital
motion  caused  by  a  StMBH.   Some  systems  in  the  9th  Catalogue
\citep{SB9}  have  high values  of  the  mass  function,  indicating a
possible  StMBH  companion,  but  they  are all  of  lower  grades  of
reliability, with  the exception of the well-studied  X-ray binary HIP
18350 = X Per. More recent discoveries include candidates, such as the
SB1  star AS  386, with  $P =  131$ days,  $K_1$  of  52 km~s$^{-1}$,  and mass
function of 1.9 $M_\odot$ \citep{Khohlov2018}. A candidate non-interacting 
system with a RG component was reported by \citet{tho}, and possibly more will
be identified in the extensive APOGEE survey \citep{bad}.

\item
Precision  photometry of  eclipsing binaries  reveals the  presence of
massive companions  via the eclipse  time variation effect. The
{\it   Kepler}    main   mission   provided    sufficient   data   for the
characterization of 222 triple systems \citep{Borkovits2016}. A few of
those  have high mass  functions, suggesting  a massive  but invisible
tertiary  companion.    However,  the  accuracy  of   this  method  is
compromised  by the  large  uncertainty of  the  period for  long-term
effects and the  possible interference from persistent, differentially
rotating photospheric spots.
\end{itemize}

\begin{deluxetable*}{cc c c c  l } 
\tablecaption{List of candidates \label{tab:list}}
\tablewidth{0pt}                                   
\tablehead{    
\colhead{TYC} & 
\colhead{HD} & 
\colhead{$V$} & 
\colhead{Spectral} & 
\colhead{$\overline{\omega}$\tablenotemark{b}} & 
\colhead{Note} \\
& &
\colhead{(mag)} & 
\colhead{type\tablenotemark{a}} & 
\colhead{(mas)} &  
}
\startdata
7381-433-1   & 318347 & 10.02 & G0 & 0.729(0.052) & Emission star \\
7390-1610-1  & 324668 & 9.71  & K0 & 1.009(0.042) &  Constant RV \\
9299-1080-1  & \ldots    & 9.74  & ?  & 1.232(0.025)  & SB1, 81 day \\
6948-350-1   & 206092 & 9.76  & G9III & 2.592(0.046)  & SB2, 4.37 days  
\enddata
\tablenotetext{a}{As given in Simbad} 
\tablenotetext{b}{Parallaxes and errors are 
  from the {\it Gaia} DR2 \citep{Gaia}.}
\end{deluxetable*}

In  this work,  we  are exploiting  the  second method,  i.e., the  RV
measurements of  the reflex orbital  motion. Our targets  are selected
from the extensive RV survey of  southern sky red giants that had been
suggested   to  serve  as   reference  stars   for  the {\it Space Interferometry Mission}
(SIM)
\citep{Makarov2015}.    The   goal  of   this   survey   was  to   vet
spectroscopically single  stars, but a large fraction  of binaries was
discovered. Only three to four individual  observations were typically  made of
each star in a campaign lasting 753 days; the sample size is 1134. 

We  selected from  the  above survey  four  stars with  very large  RV
variation for further monitoring, with  the aim to confirm their large
amplitudes  and determine  the orbits.   A large  minimum mass  of the
secondary component derived from the spectroscopic orbit would provide
a strong indication that the companion could be a StMBH and the object
is  a red  giant  (RG)+BH binary.  The  periods of  such binaries  are
expected to be longer than $\sim$100  days owing to the large radii of
giants. Binaries  with shorter orbital  periods should go  through the
common envelope  stage and end up  as merged stars (prior  to the core
collapse) or as tight  low-mass X-ray binaries \citep{iva}. At
longer periods than $\sim$100 days, a giant star of 2 $M_\odot$ orbited by a 5
$M_\odot$ BH would  have an RV amplitude of $K_1  = 60$ km~s$^{-1}$ if
the  orbit is  seen edge-on.   At  longer periods,  $P$, the  amplitude
decreases as $P^{-1/3}$.  A 2 yr  SIM grid survey could reveal RG+BH
binaries with periods up to $\sim$5 yr.

\begin{figure}
\plotone{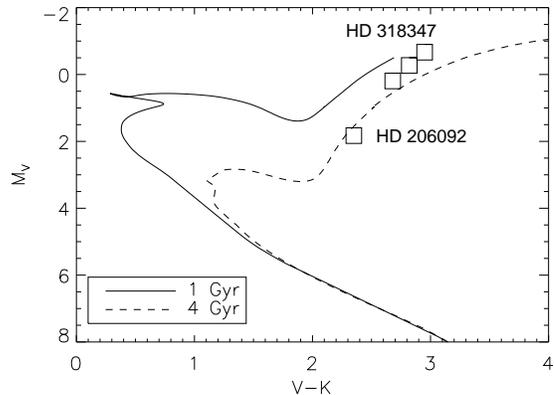}
\caption{Location of the four candidates on the $M_V, V-K$
  color-magnitude diagram. The lines are isochrones for solar
  metallicity and ages of 1 and 4 Gyr from \citet{Dotter2008}. De-reddening corrections are not applied.
\label{fig:CMD} }
\end{figure}

None of our targets turned out to contain spectroscopically detectable
massive companions.  Nevertheless, the  claimed large RV variation had
to be verified and explained, revealing some 
intrinsically  interesting and rare stars.   The   targets  and  our  observing   method  are  briefly
introduced      in      Section~\ref{sec:obs}.      The      following
Section~\ref{sec:obj}  presents our results  regarding each  star. The
general     discussion     and     conclusions    are     given     in
Section~\ref{sec:disc}.

%%--------------------------------------------------------------
\section{Observational data}
\label{sec:obs}

\subsection{Targets}

Four stars  with large RV variations  were selected from  the SIM grid
survey.   Table~\ref{tab:list}  provides basic  data  on these  stars:
their  common  identifiers  (they  allow  for the  retrieval  of  other
information   from  Simbad),   visual   magnitudes,  spectral   types,
parallaxes, and  short notes. Figure~\ref{fig:CMD}  shows the location
of  these stars  on the  $(M_V, V-K)$  color-magnitude  diagram (CMD),
computed  using {\it  Gaia} DR2  parallaxes \citep{Gaia}  and  the $K$
magnitudes given  by Simbad. No de-reddening  corrections are applied.
The stars are  elevated above the main sequence,  confirming the giant
status.

\subsection{CORALIE RVs}

The  RV  survey of  SIM  grid stars  was  conducted  by D.~Queloz  and
D.~S\'egransan using  the CORALIE echelle spectrometer  at the 1.2 m
Euler   telescope  in   La  Silla.    The  RVs   were   determined  by
cross-correlating   the   reduced  spectra   with   the  binary   mask
\citep{CORALIE}.  The cross-correlation  function (CCF) contains a dip
produced by  all of the absorption lines  included in the mask.  Its position
defines  the RV,  while the  width of the  dip depends  on the width  of the
stellar lines (hence on the projected rotation velocity). The analysis
by \citet{Makarov2015} shows the high overall quality of these RVs: the
mean absolute error is 34 m~s$^{-1}$, and the intrinsic RV jitter caused by the
atmospheres of red giants is of the same order. 

Observations reported  here revealed that the large  RV variability of
our candidates  was, mostly,  caused by occasional erroneous measurements in  the CORALIE
data.  As noted by D.~S\'egransan (2018, private communication), the
CCF  may not  contain  a valid  dip  for several  reasons: very  noisy
spectrum (e.g. taken through  the clouds), absence of absorption lines
matching the  mask, e.g.  a  star of early  spectral type, or a
large RV that falls outside the normally computed CCF window.  In such
cases, the processing software measures a wrong RV using the local CCF
minimum. Outlying measurements could be  identified and rejected by the low
dip  contrast. Problematic detections   are    further    discussed    in
Section~\ref{sec:obj}.

\begin{figure}
\plotone{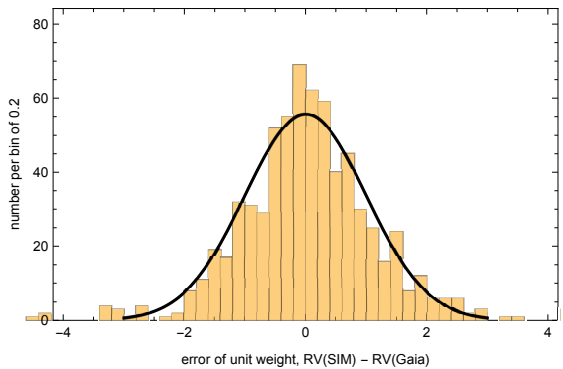}
\caption{Histogram of normalized RV differences between the SIM RV survey and Gaia DR2 catalog
for 697 common stars with binarity probability less than 0.9. The thick line shows
the Normal$[0,1]$ distribution for comparison. A common zero-point shift of $+0.27$ \kms
was applied to all RV(SIM) $-$ RV(Gaia) differences. The core of the empirical distribution
is narrower than the expected distribution, indicating that the standard errors of mean RV
may be slightly overestimated.
\label{dr2.fig} }
\end{figure}

To confirm the overall high quality and variability of the SIM RV survey, we compared
the data with the mean RVs from the Gaia Data Release 2 (DR2). We cross-matched
1133 stars in DR2, but only 1084 have RV measurements there. Selecting only
stars with a probability of binarity (given in SIM RV) below 0.9, the resulting
sample counts 697 stars. Fig. \ref{dr2.fig} shows the histogram of the RV unit
weight error, i.e., the observed difference of RV(SIM) and RV(DR2) divided by
the quadratic sum of their errors provided in both catalogs. The theoretically
expected Normal$[0,1]$ PDF is shown for reference. In order to make the center
of the empirical distribution coincide with 0, we added a common zero-point
shift of $0.27$ \kms to all individual differences. The presence of a systematic bias
in DR2 measurements was discussed by \citet{kat}. Fig. \ref{dr2.fig} confirms
that the bulk of RV measurements of constant stars are as precise as their standard errors suggest
in both catalogs.

\subsection{CHIRON observations}

The observations  reported here were conducted at  the 1.5 m telescope
located at Cerro Tololo (Chile) and operated by the SMARTS consortium.
Ten hours of observing time  were allocated through the National Optical
Astronomy Observatory (NOAO). Spectra were
taken  by the  telescope operator  in the  service mode.   The optical
echelle spectrometer CHIRON \citep{CHIRON}  was used in the fiber mode
with a spectral resolution of 30000.  On each visit, a single 5-minute
exposure of  the star  was taken, accompanied  by the spectrum  of the
comparison lamp for wavelength  calibration.  The data were reduced by
the pipeline written in IDL.

The RVs are derived from the reduced spectra by cross-correlation with
a binary  mask based on the  solar spectrum, similarly  to the CORALIE
RVs. We  used  39 echelle  orders  in the  spectral range from
4500 to  6500\AA, which is relatively less contaminated by telluric
lines.  More details are  provided by \citet{Tok2016}. The RVs delivered by this procedure should be on
the  absolute  scale  if   the  wavelength  calibration  is  good.   A
comparison of  CHIRON RVs with  several RV standards revealed  a small
offset  of $+0.16$  km~s$^{-1}$ \citep{Tok2018};  in the  following this
offset is  neglected. The mean  RVs of the  star with constant  RV, HD
324668, measured by CORALIE and  CHIRON, differ only by 43 m~s$^{-1}$.

Table~\ref{tab:orb}  gives the  elements of  two  spectroscopic orbits
derived from the CHIRON RVs. The  notations are standard, as is the method
of orbit calculation by weighted least squares. Individual RVs are
given in Table~\ref{tab:rv}, published in full electronically.

\begin{deluxetable*}{l ccc c ccc c c  }
\tablecaption{Orbital elements \label{tab:orb}}
\tablewidth{0pt}                                   
\tablehead{    
\colhead{Name} & 
\colhead{$P$} & 
\colhead{$T_0$} & 
\colhead{$e$} & 
\colhead{$\omega$} & 
\colhead{$K_1$ } & 
\colhead{ $K_2$} & 
\colhead{ $\gamma$} & 
\colhead{$N$} & 
\colhead{rms} \\ 
&
\colhead{(day)} & 
\colhead{-2400000} & 
&
\colhead{(deg)} & 
\colhead{\kms} & 
\colhead{\kms} &  
\colhead{\kms} & &
\colhead{\kms} 
}
\startdata
 TYC 9299-1080-1 &  80.99 & 58394.76          & 0          & 0       & 18.41   & \ldots     & 317.71      & 13 & 0.37  \\
       & $\pm$0.02 & $\pm$0.08                & fixed      & fixed   & $\pm$0.13 & \ldots   & $\pm$0.08    & \ldots   & \ldots \\ 
HD 206092 &  4.37545& 58271.347         & 0          & 0    & 43.88  & 111.69     & 27.35   & 17 & 1.96  \\
       & $\pm$0.00008 & $\pm$0.008 & fixed      & fixed   & $\pm$0.87 & $\pm$1.05   & $\pm$0.50  & \ldots   & 1.55    
\enddata
\end{deluxetable*}

\begin{deluxetable}{l ccc}
\tablecaption{Radial velocities \label{tab:rv}}
\tablewidth{0pt}                                   
\tablehead{ 
\colhead{Name} & 
\colhead{JD} & 
\colhead{RV} & 
\colhead{Comp.} \\
& 
\colhead{$-2400000$} &
\colhead{\kms} & 
}
\startdata
HD 324668 &  58260.8470 &-32.651 & \\
HD 324668 &  58270.7916 &-32.624 & \\
HD 324668 &  58341.5915 &-32.590 & \\
 HD 206092 & 58260.912 &   -6.810 & a \\
 HD 206092 & 58260.912 &  112.381 & b \\
 HD 206092 & 58271.899 &   59.516 & a \\
 HD 206092 & 58271.899 &  -50.498 & b
 \enddata
\end{deluxetable}

\subsection{Speckle interferometry}

All  targets were observed  on 2018.4  in the  $I$ band  using the speckle
camera at  the 4.1 m SOAR telescope.  The  angular resolution (minimum
detectable  separation) was  50\,mas, and  the dynamic  range (maximum
magnitude difference)  is about 4  mag at 0\farcs15  separation.  The
instrument and  observing technique are described  in \citet{SOAR}. No
companions were detected.

%%--------------------------------------------------------------
\section{Notes on individual objects}
\label{sec:obj}

\subsection{HD 318347}

Hydrogen emission in  the spectrum of this star has  been noted a long
time ago. Simbad gives the spectral type  G0, matching the red color $V -K
= 2.96$ mag. However, the star  is featured in the catalog of galactic
OB stars by \citet{Reed2003}. 

\begin{figure}
\plotone{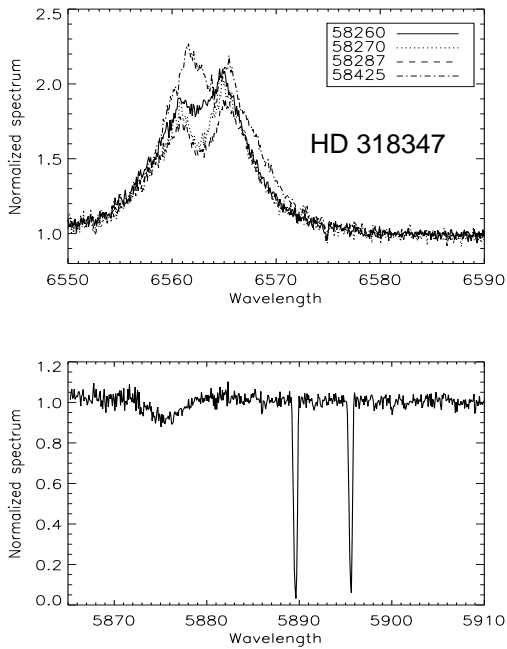}
\caption{Two echelle orders in the spectra of  HD~318347 
containing the  H$\alpha$ line  (top) and  the sodium D lines
(bottom), normalized by the blaze function. Note the  strong
variability of the H$\alpha$ emission line strength in the 
spectra taken on four different MJD epochs. 
\label{fig:HD318347} }
\end{figure}

The four  CHIRON spectra taken  over 165 days  (from JD 2458260  to JD
2458425)  do not  have absorption  lines typical  of  late-type stars,
apart from  the strong sodium absorptions,  apparently of interstellar
origin.  The  H$\alpha$ line shows a  strong and wide  emission with a
double peak  (Fig.~\ref{fig:HD318347}).  The strength of  the emission and
the  contrast  of  the  two  peaks  change on  the  time  scale  of  a
fortnight. The SIM  grid catalog lists six RVs  ranging from $-245$ to
$323$   km~s$^{-1}$,  all   with   small  errors   not  exceeding   25
m~s$^{-1}$. Suspiciously,  two RVs measured  with CORALIE on  the same
night, JD 2453591, differ by  54 km~s$^{-1}$.  These RVs were probably
derived from the CCFs without valid dips, given the lack of absorption
lines in the spectrum.

Most likely, HD~318347 is a highly reddened early-type star located in
the  Galactic plane  at a  distance of  1370 pc.  However, \citet{hou}
find, based on  LAMOST data, that most of  the double-peaked H$\alpha$
emission  line stars  appear in  binaries. A  short-period cataclysmic
variable  cannot be precluded.  It is  possible that  the unidentified
{\it  Fermi} Large  Area  Telescope (LAT)  source 2FGL  J$1746.5-3228$
\citep{nol}  is   associated  with   the  {\it  Swift}   X-ray  source
J$174645.4-323746$  \citep{pag} and with  HD~318347, which  is located
$5.8$  arcsec  away,  a  little  more than  the  estimated  positional
uncertainty of the former.  The 10\AA ~width of the H$\alpha$ emission
implies gas motions at $\sim500$ km~s$^{-1}$, possibly associated with
accretion onto  the stellar surface. Circumstellar  material can cause
additional extinction as well as the infrared excess.

\subsection{HD 324668}

This K0 giant is located in the Galactic plane at a distance of
1~kpc. The average of the three CHIRON RVs is $-32.622$ km~s$^{-1}$ with an rms scatter
of 31 m~s$^{-1}$. They match perfectly the two RVs measured by CORALIE, with
an average of $-32.665$ km~s$^{-1}$. However, the third CORALIE RV 
of $+276.3$ km~s$^{-1}$ is highly discrepant, earning this star a title of
spectroscopic binary in Simbad. We can only guess whether this
discrepancy was caused by pointing at another star in this crowded sky
region or for some other reason. Obviously, this red giant has a constant
RV. Incidentally, it proves  the excellent agreement between the RV zero
points of CORALIE and CHIRON.

\subsection{TYC 9299-1080-1}

\begin{figure}
\plotone{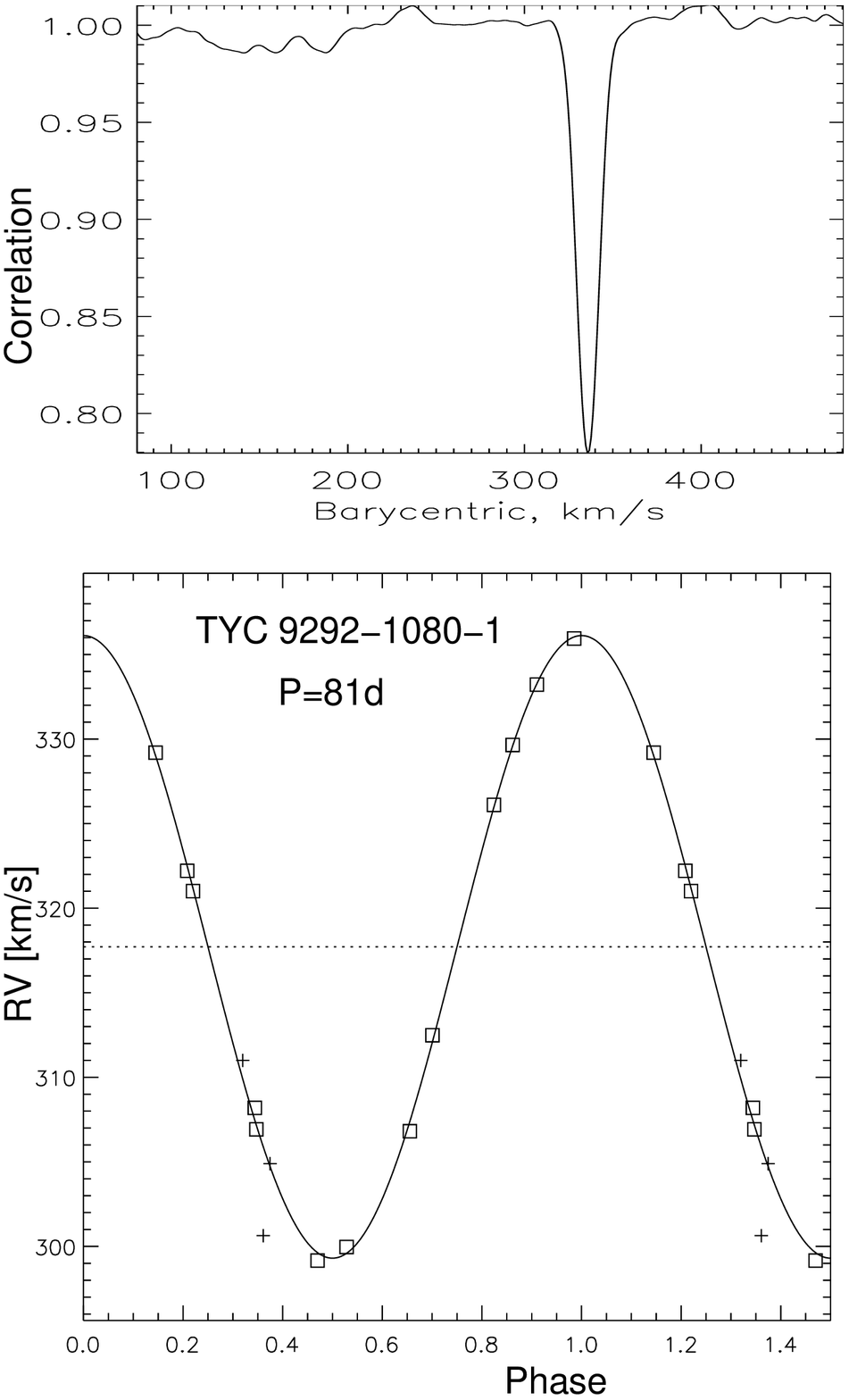}
\caption{The CCF (top) and RV curve (bottom) of 
	TYC 9299-1080-1. The crosses mark two RVs from CORALIE and one
        from {\it Gaia}.
\label{fig:TYC} }
\end{figure}

This star,  also known as  CD$-$72 1472, is  located at a  distance of
812~pc and has a Galactic  latitude of $-25^\circ$.  All three CORALIE
RVs are mutually discordant: $+300.6$, $-273.9$, and $+552.6$ km~s$^{-1}$.
They were measured over a time  span of 696 days. The first CHIRON RVs
have shown  a slow  trend, inspiring  hope that this  star has  a long
period and  a large RV amplitude.  However,  further observations have
shown that  the RV varies with a  period of 81 days (Fig. \ref{fig:TYC}). Elements of the
circular single-lined  orbit derived from 13 CHIRON  RVs are presented
in Table~\ref{tab:orb}. The {\it  Gaia} median RV of $311.0$ km~s$^{-1}$
roughly  matches   our  orbit,  assuming  an  epoch   of  2015.5.   The
variability  of  the Gaia  RV  detections  can  be inferred  from  the
elevated  error of  the median  ($5.7$ km  s$^{-1}$), which  implies a
single-measurement standard  deviation of  $\sim 19$ km  s$^{-1}$. The
first  CORALIE  RV  fits  the orbit  crudely.   D.~S\'egransan  (2018,
private communication)  provided another RV measured by  CORALIE on JD
2454699.52, which  matches the orbit  perfectly. We attribute  the two
discrepant CORALIE RVs to the large  RV of this star, placing the true
dip outside the nominal CCF window. This explains why this object
has  two wrong  RVs.  However,  the small  errors  of these  erroneous
measurements reported  by the CORALIE pipeline, 22  and 16 m~s$^{-1}$,
are perplexing.

For the mass of the primary star in the range from 1 to 2 $M_\odot$,
the minimum mass of the companion is from 0.48 to 0.73 $M_\odot$. The companion could be
a normal  solar-type dwarf or a  white dwarf. The large  radius of the
giant  primary has  caused tidal  orbit circularization.   Its unusual
feature  is the  large center-of-mass  RV of  317.7  km~s$^{-1}$.  The
parallax and  proper motion correspond  to the tangential  velocity of
217 km~s$^{-1}$,  hence the  star moves with  a total velocity  of 384
km~s$^{-1}$ relative  to the  Sun.  Considering the  Galactic rotation
velocity at  the solar radius,   236$\pm$3 \kms  \citep{Kawata2019}, and
the    solar    peculiar    velocity    $(U_\odot,V_\odot,W_\odot)    =
(6.0,10.6,6.5)$ \kms \citep{Bobylev2014}, we derive the galactocentric
velocity of TYC 9299-1080-1, $(U,V,W)_{\rm gal} = (302.9, 10.0, 66.2)$
\kms.   This binary star  certainly belongs  to  the Galactic  halo. It  moves
almost  straight toward  the  galactic  center and  will  pass in  its
vicinity. However, the motion is not fast enough for a runaway or extragalactic star. Our numerical
integration of the Galactic orbit indicates that it moves on a highly extended orbit between
$\sim100$ pc and 20 kpc from the center, which will be reached in less than 20 Myr.

%I applied the Galactic rotation at solar radius of 236 +-3 km/s from Kawata, D. et al. 2019, MNRAS 482, 40 and the solar peculiar velocity (u_\odot,v_odot,W_\odot) = 6.0,10.6,6.5 km/s from Bobylev, V.V., Batkova, A.T., 2014, MNRAS 441, 142 (alternatively, I could use the value from my earlier paper with D. Murphy, which is close). The result is (U,V,W) galactocentric = (302.9, 10.0, 66.2) km/s. I also integrated its Galactic orbit, if that would be of any interest. It is certainly a halo star, moves almost straight to the center, and passes close to the Galactic center. But it is not runaway, being bound to the Galaxy. The runaway velocity would be somewhere above 500 km/s.

\subsection{HD 206092}

\begin{figure}
\plotone{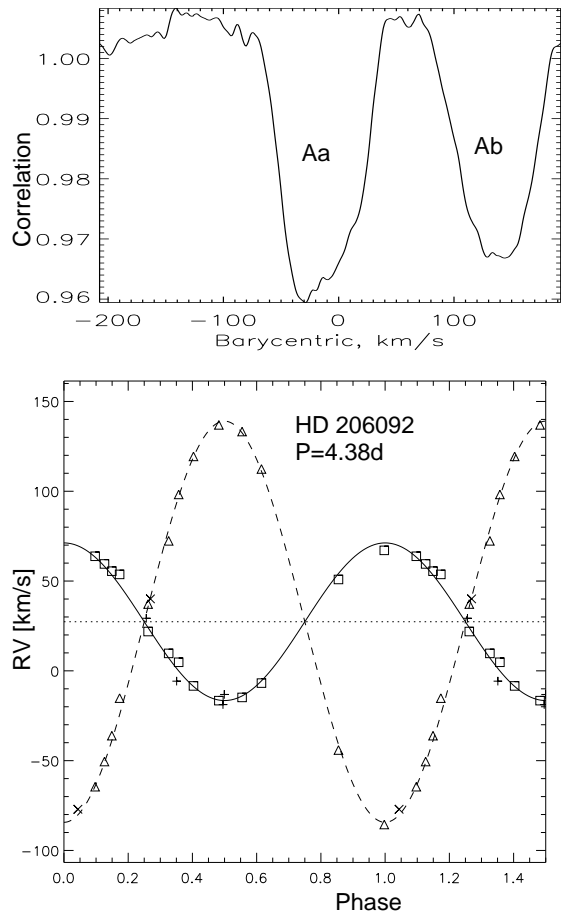}
\caption{The CCF on JD 2458356 (top) and the RV curve (bottom) of 
HD 206092. The squares and full line denote the primary component Aa,
the triangles and dashed line correspond to the secondary Ab, and the crosses and
plus signs depict the CORALIE RVs.
\label{fig:HD206092} }
\end{figure}

Located at a distance of 386 pc, this object is the closest of our
four candidates. Its association with a ROSAT X-ray source was noted, while
\citet{Kiraga2012} found photometric variability with a period of
2.1876 days, presumably caused by rotational modulation. 

The first  CHIRON spectra revealed  wide double lines in  rapid motion
(Fig.~\ref{fig:HD206092}).    The  system's   location   in  the   CMD
(Fig.~\ref{fig:CMD}) indicates  that the  primary is an  evolved star,
probably a red giant. Thus, this system belongs to the rare type of RS
CVn-type binaries  characterized by  high levels of  chromospheric and
X-ray activity.  Further monitoring  established the orbital period of
4.38  days.    The  orbit  (Table~\ref{tab:orb})   is  circular,  with
amplitudes of 44 and 112  km~s$^{-1}$.  The residuals of CHIRON RVs to
the orbit are relatively large because  the CCF dips are wide and have
somewhat  irregular  shapes,  presumably  caused  by  starspots.   The
photometric period found by  \citet{Kiraga2012} corresponds to half of
the  orbital  cycle,  indicating  ellipsoidal  variability  caused  by
tidally distorted stars in a close binary. The Balmer lines H$\alpha$
and H$\beta$ in the spectra of  HD 206092 are shallow, being filled by
emission;  the sodium  D  lines  also have  emission  in their  cores
produced in the chromosphere. 

In  this  case,  the  large  RV variability  detected  by  CORALIE  is
explained by  the orbital motion. We  compared these six  RVs with our
orbit and found that one is highly discrepant, three correspond to the
primary component Aa  and two match the secondary  Ab.  These RVs were
used with  low weights to improve  the accuracy of the  period. The RV
errors reported by CORALIE for this star are unrealistically small.

The orbit  corresponds to the minimum  masses $M_1 \sin^3  i$ and $M_2
\sin^3 i$ of 1.22 and 0.48  $M_\odot$, or the mass ratio of 0.39. Yet,
the areas of  the CCF dips are comparable,  indicating the light ratio
of  0.82 (see Fig.~\ref{fig:HD206092}).   Assuming the  mass sum  of 2
$M_\odot$, the orbital  axis is 14.2 $R_\odot$. The  wide CCFs suggest
that the components almost touch  each other; the large luminosity and
the red color correspond to large stellar radii. The roughly estimated $v\sin i$
(projected surface velocity of rotation) is 40 \kms, hence, the radii for
both components $R>3.5 R_\sun$ assuming synchronous rotation. The unusually bright
(for its mass ratio) secondary  component Ab must be transferring mass
to the primary  component Aa. We tried to find  analogs of HD 206092
by  selecting binaries  with  the following  parameters: period  $<10$
days, $V-K>2$  mag and $M_V < 2.5$  mag.  The only analog  we found is
RS~CVn  (HD  114519), the  prototype  of  its  class, which is defined as chromospherically
active, short-period binaries with primaries evolved off the main sequence
\citep{fek}. The  corresponding
parameters for RS  CVn are $P=4.80$ days, $M_V$  = 2.5 mag, $V-K=2.16$
mag, and spectral  type K0IV. Not  surprisingly, the object is  a powerful
source of coronal X-ray emission listed in the ROSAT catalog of bright
sources.

Looking at evolved binary systems with somewhat longer periods, the large mass
ratio and similar line strengths of HD 206092 suggest that
it is similar to chromospherically active semi-detached
eclipsing binaries such as AR Mon and RZ Cnc \citep{pop}. Although AR Mon and RZ Cnc have significantly
longer periods (20 days) than HD 206092, they appear to be in a similar
mass transfer evolutionary stage. Another similar large mass
ratio system, which has a shorter period of 10.7 days, is
RV Lib \citep{imb}.

Using a ROSAT-determined count rate of CR$=0.176(0.023)$ cts s$^{-1}$ and a hardness
ratio of HR1$=0.44(0.13)$ we derive an X-ray luminosity of $L_X=3.34(0.066)\times
10^{31}$ erg s$^{-1}$, where the standard error is estimated from the formal
uncertainties of the CR, HR1, and Gaia parallax. RS CVn-type binaries
are the most luminous (and, on average, the hardest) sources of coronal
X-ray radiation \citep{mak}, but HD 206092 is almost twice as luminous as the
brightest X-ray star within 50 pc of the Sun, which is II Peg, also a RS CVn-type
binary. Another useful comparison is the prototype star RS CVn (F6IV+G8IV) with
$L_X=1.67(0.01)\times10^{31}$ erg s$^{-1}$. This puts HD 206092 into the
category of outstandingly active and X-ray luminous field stars.

\section{Discussion and conclusions}
\label{sec:disc}

\subsection{Expected number of RG+BH binaries}

In the past, considerable effort  has been spent to predict the number
of binaries  containing compact objects, focusing  mostly on potential
sources of gravitational waves  generated by merging binaries with two
compact  components \citep[e.g.][]{B2002}.   These  works use
binary  population  synthesis.  The  results  depend  strongly on  the
initial assumptions such as binary  statistics and are also affected by the
uncertainties  in   the  binary  and  stellar   evolution.   The  same
population synthesis approach  was used more recently  to estimate the
number  of   star-BH  binaries   detectable  by  {\it   Gaia}  through
astrometric   effects    of an unseen   companion    on   the   visible
star. Predictions  made by  different groups differ  by two  orders of
magnitude,   depending   on    the   assumptions   and   models   used
\citep{Kinugawa2018}. 

All  massive binaries  with orbital  separations less  than 20 au are
affected by the mass transfer  and mass loss that changes their orbits
\citep{Moe2017}.   Large velocities  acquired by  the  remnants during
supernova  explosions  (kicks),  on   the  order  of  hundreds  of
km~s$^{-1}$,   likely  destroy  all   binaries  except   the  tightest
ones. Hence,  it is possible  that RG+BH binaries with  periods longer
than $\sim$100 days do not exist\footnote{Several {\it low-mass} X-ray binaries with giant
components have been identified, however. For example, symbiotic X-ray
binaries are a rare class of low-mass, hard X-ray binaries
that consist of a neutron star accreting mass from an M giant.
Two systems so far have been analyzed, V2116 Oph \citep{hin6}
 and V934 Her \citep{hin18} and both have long periods of 3.2 and 12.0
years, respectively. Thus, at least for supernova remnant
binaries that result in neutron stars, some long-period systems
do survive.}.

Neglecting both orbital evolution  and kicks, we estimate the fraction
of RG+BH progenitors,  i.e.  the upper limit of  the fraction of RG+BH
binaries  among red  giants.  The  progenitors of  BH  components have
masses of $M_{0} >  20  M_\odot$ \citep{B2002},  while  the giants  have
typical masses between $M_1 = 2 M_\odot$ and $M_2 = 3 M_\odot$.  Using
the Salpeter  mass function, $f(M) \propto M^{-\alpha}$  with $\alpha =
2.35$,  we evaluate  the fraction  of  stars more  massive than  $M_0$
relative to stars with masses  between $M_1$ and $M_2$, $f_* = 0.106$.
Now, some  fraction of  stars with $M_{0}  > 20 M_\odot$  are binaries
with  secondary components  in  the  same mass  range  as our  giants,
between $M_1$  and $M_2$,  and with periods  from 100 to  $10^3$ days.
These binaries  are potential progenitors of RG+BH  objects when their
primary components  become BHs  and the secondaries  turn into
giants.  The fraction  of such progenitors relative to  field stars in
the same mass range (which  also become giants) can be estimated using
the recent  analysis of  binary statistics by  \citet{Moe2017}.  About
$f_B=0.2$ of massive stars have binary companions of all masses in the
above period  range. At  those periods, the  distribution of  the mass
ratio $q$  does not follow  the Salpeter function, although  it still
grows  as $q^{-1.5}$ at  $q>0.3$. Therefore,  the fraction  of massive
stars with suitable  parameters is $f_B f_q$, where  $f_q \approx 0.2$
is the fraction of companions in the selected mass range between $M_1$
and $M_2$.  Summarizing for each primary star in  this mass range we
get $f_* f_B f_q = 0.004$ potential progenitor binaries.

A  sample of  1000  giants is  not  expected to  contain  more than four
progenitors of  RG+BH binaries  with intermediate periods.   Given the
evolution  of  binary  orbits  and  the  destructive  effects  of  the
supernova kicks, the expected number  of RG+BH binaries should be much
less than the number of their progenitors. Hence, the non-detection of
such objects in the SIM grid survey is natural.

\subsection{Related work}

Recently, \citet{Murphy2018}  introduced a new  observing technique by
deriving  quasi-spectroscopic  orbits  from   the  timing   of  stellar
pulsations. They  presented a  sample of 314  such orbits for primary
stars of  A/F spectral  types and periods  between 100 and  1500 days,
progenitors of  the red  giants studied here.  The size of  the parent
sample  was  2224,  twice as big as the  SIM  grid  sample.   They
estimated that  a fifth of these binaries  have degenerate white dwarf
secondaries, while their primaries are products of mass transfer (blue
stragglers).   None  of  the   binaries  had  a  large  mass  function
indicative of the BH secondary. This non-detection of BH secondaries
agrees with our result. 

Using population synthesis,  \citet{Kinugawa2018} estimated the number
of  binaries with  BH  components detectable  astrometrically by  {\it
  Gaia}. Their assumptions regarding binary statistics differ from the
latest analysis by \citet{Moe2017} in several important respects.  The
predicted fraction of  BH binaries with periods between  50 days and 5
yr  (similar to  the period  range  explored here)  in a  simulated
sample  of $10^5$  progenitor  binaries ranges  from  0.013 to  0.028,
depending  on   the  metallicity  (more  BH   binaries  in  metal-poor
population). The majority of these binaries contain stars less massive
than 2 $M_\odot$. We stress that these estimates are highly uncertain.

\subsection{Conclusions}

The large RV variability of some red giants detected by the CORALIE
survey of SIM grid stars was intriguing and called for further
investigation. We found that this detection is spurious, resulting
from flukes in data reductions. 

Theoretical estimates, still highly uncertain, indicate that the
fraction of red giants containing an StMBH remnant cannot exceed
$10^{-2}$, and likely is orders of magnitude smaller. Yet,
observational limits on the existence of such RG+BH binaries should be
placed independently of the theory. Our study contributes to
establishing such limits by non-detection of RG+BH candidates in a
sample of 1000 stars. 

Among  the four  candidates studied  here, two  can be  of independent
interest: the  red-giant binary TYC 9299-1080-1 with  a large spatial
velocity of 384 km~s$^{-1}$  and the semi-detached binary HD~206092 of the rare
RS CVn type.

\section*{Acknowledgments}

We thank the telescope operator R.~Hinohosa for taking the data. 
D.~S\'egransan has kindly helped us to understand the origin of
discrepant CORALIE RVs and provided additional unpublished RV
of TYC 9299-1080-1.

%%--------------------------------------------------------------
%\section{}
%\label{sec:}

%%--------------------------------------------------------------
%\section{}
%\label{sec:}

%%--------------------------------------------------------------
%\section{}
%\label{sec:}

\end{document}